\documentclass[10pt]{article}

\usepackage{epsfig}

\newlength{\dinwidth}
\newlength{\dinmargin}
\marginparpush
5pt \topmargin -42pt \headheight 12pt

\begin{document}
\titlepage

\vspace*{4cm}

\begin{center}

{\Large \bf May Heavy neutrinos solve underground and cosmic ray
puzzles?}
%\title{Observational indications in favor of primordial heavy neutrino
%existence }

\vspace*{1cm}
\textsc{K. Belotsky$^{a,b}$, D. Fargion$^{c,d}$,
M. Khlopov$^{a,b,c}$, R.V. Konoplich$^{c,e}$} \\

\vspace*{0.5cm}
$^a$ Moscow Engineering Physics Institute,
Moscow, Russia \\[0.5ex]
$^b$ Center for Cosmoparticle Physics "Cosmion" of
Keldysh Institute of Applied Mathematics, \\
Moscow, Russia \\[0.5ex]
$^c$ Universita' di Roma "La Sapienza" and INFN, Rome, Italy \\[0.5ex]
$^d$ INFN Section Roma1, Rome, Italy \\[0.5ex]
$^e$ Department of Physics, New York University,\\
New York, NY 10003, USA\\
Department of Physics, Manhattan College, Riverdale, New
York, NY 10471, USA

\end{center}

\vspace*{1cm}

\begin{abstract}

Primordial Heavy neutrinos of 4th generation might explain
different astrophysical puzzles : indeed the simplest 4th neutrino
 scenario may be still consistent with known 4th neutrino
physics, cosmic ray anti-matter and gamma fluxes and signals in
underground detectors for a very narrow neutrino mass windows
($46-47$ GeV). We have analyzed extended Heavy neutrino models
related to the clumpiness  of  neutrino density, new interactions
in Heavy neutrino annihilation, neutrino asymmetry,  neutrino
decay. We found that in these models the underground signals maybe
better combined with the cosmic ray imprint leading to a wider
windows for neutrino mass  ($46-75$ GeV) coinciding with the
whole range allowed  from uncertainties of electro-weak
parameters.
\end{abstract}

%\newpage

\section{Introduction}

The problem of dark matter (DM) of the Universe was revealed about
70 years ago. Several possible physical candidates were suggested
since that time and several approaches to probe these candidates
appeared. However, observed phenomena, which have unclear nature
yet, could not be decisively matched with expected effects
caused by existing candidates.

An important step in exploration of DM problem was a development
of direct searches for Weakly Interacting Massive Particles
(WIMP). Underground detectors were created in which effects of
nucleus recoil induced by interaction of cosmic WIMP with nucleus
were searched for. A positive result at 6.3 sigma C.L. was obtained
in the DAMA/NaI underground set-up at the Gran Sasso National Laboratory of
I.N.F.N.
by exploiting the distinctive WIMP annual modulation
signature \cite{DAMA-review}. Being model independent
this positive result cannot be directly compared with the single
model-dependent negative results
of other groups, which have also used different target-nuclei, different
experimental strategies, different set-ups
and all assumptions fixed at a single set  \cite{DAMA-review}. Moreover, it
can be shown  \cite{DAMA-review} that
these negative results are actually not incompatible with the positive
signal by DAMA/NaI.

On the other hand, an indirect probing of DM can be based on
cosmic ray (CR) data. The presence of DM in Galaxy in form of
WIMPs can cause an appearance of cosmic particles of high energy
due to an annihilation or decay of the WIMPs. An implication of
these annihilation (decay) sources of cosmic rays could remove
possible contradiction between observed cosmic ray fluxes and
their predictions on the base of standard cosmic ray model.

The result of DAMA/NaI is a challenge for DM studies, expected to shed
a light on the origin of DM. In fact, the existing physical
candidates of dominant DM can hardly or not at all provide an
explanation of the result of DAMA/NaI. WIMP candidates such as
neutralino, axion, gravitino, sterile neutrino, axino, mirror
(shadow) matter are able to compose all the required missing mass
of the Universe, however, all these candidates, except neutralino,
are virtually sterile particles in respect to their interactions
with an ordinary matter. Therefore, it looks like the measurements
of DAMA/NaI as well as anomalies observed in cosmic rays spectra
require a non-sterile DM which, in particular, could be a
non-dominant DM component in the form of Heavy neutrinos of the
4th generation \cite{Fargion}.

A possibility to explain the DAMA/NaI result within the framework of
Standard Model extended to the 4th generation of fermions,
revealed in \cite{Fargion}, is the subject of current
consideration.

The Heavy neutrino ($N$) is supposed to be a neutral fermion of
a new 4th generation possessing the standard weak interaction.
According to recent analysis of precision electroweak
data\cite{Okun}, where possible virtual contributions of 4th
generation particles were taken into account, a fit is compatible
with the 4th neutrino, being Dirac and (quasi-)stable, in a mass
range about 50 GeV (47-50 GeV is $1\sigma$ interval, 46.3-75 GeV
is $2\sigma$ interval) \cite{Okun} and other 4th generation
particles satisfying their direct experimental constraints (above
80-130 GeV). In the following we will assume that the 4th neutrino
mass is about 50 GeV.

If the fourth neutrino is sufficiently long living or absolutely
stable, its primordial gas from the early Universe can survive to
the present time and concentrate in the Galaxy. In the case of
charge symmetry of 4th generation particles the primordial 4th
neutrinos can not account for a bulk of missing mass in the
present Universe. In the mass range about $50-80$ GeV the 4th neutrinos
can make up $10^{-5}-10^{-2}$ of total density of the Universe,
being a non-dominant DM component. This leads to a scenario of
multi component dark matter consisting of a subdominant Heavy
neutrino component and a sterile dominant component. A complex
analysis of astrophysical effects induced by a presence of Heavy
neutrino (non-sterile) DM component is the purpose of present
paper.

It is worth noting that the 4th generation of quarks and leptons,
considered here and neutralino, which is widely considered as the
candidate for WIMPs,  are naturally incorporated in the framework
of heterotic string phenomenology. \footnote{Mirror(shadow)
matter, also naturally follows from heterotic string
phenomenology. It is usually considered as sterile but its WIMP
effects may  play a role\cite{Foot}}. It appeals to future
multi-component dark matter analysis of the results of direct and
indirect WIMP searches. However, the astrophysical uncertainties
revealed below even for the model of 4th generation neutrino,
which is the simplest physical model and implies the minimal
amount of parameters, demonstrate all the complications to be
expected in such multi-component dark matter approach.

We should remark that even if in principle the UHE
Neutrino of fourth family might produce a very exciting
scattering with relic Heavy neutrino (a Heavy-Z-Boson Burst)
nevertheless the cosmic relic N are too diluted and poor in number to
be anyway competitive with the much more abundant (by $13$ order of
magnitude)
and effective light neutrino \cite{Z-Far}
scattering. So there is little consequence of any  Heavy-Z-Boson Burst
model interactions.
The eventuality for a UHE Neutrino (produced, for
instance in Top-Down models as \cite{Khlopov})
to be a source of amplified resonant interaction with electron or
quark (as for SUSY UHE neutralino scattering into s-electron and s-quark
channels \cite{Datta}) is absent:
there are not s-channel interaction able to overcome the electro-weak
cross-sections but only much less effective t-channel processes.
Finally, the UHE N created by Top-Down models may nevertheless
induce charged and neutral current interactions in Neutrino Detectors
(SK,UNO,$km^3$) almost  un-distinguishable from lighter UHE neutrino
scattering in matter. This effect has just the ability to increase by a small
factor ($\sim 25 \%$) the event rate in $km^3$ or EUSO neutrino induced events
if Top-Down mechanism is the main source of UHECR.

\section{Estimation of local Heavy neutrino density from DAMA/NaI
experiment}

A contribution of Heavy neutrinos $\rho_{loc\,N}$ to the total
local density $\rho_{loc}$ is given by a ratio
\begin{equation}
\xi_{loc}=\frac{\rho_{loc\,N}}{\rho_{loc}}.
\end{equation}
Approximately this parameter can be estimated as
$\Omega_N/\Omega_{CDM}$ by assuming a dominance of Cold Dark
Matter (CDM) in the Galaxy and by choosing the local fraction of relic
Heavy neutrinos equal to their contribution to the cosmological density of
CDM.

The results of DAMA/NaI, based on measurements of an "active" DM
component, i.e. in our assumptions on cosmic Heavy neutrinos, give
the fraction $\xi_{loc}$ of Heavy neutrinos in the local galactic
density. Heavy neutrinos interact with nuclei
($^{23}Na,\,^{127}I$) of DAMA/NaI detector through the
spin-independent coherent vector weak coupling. A spin-dependent
axial weak coupling of neutrino and nuclei would contribute
significantly within the considered mass range only if
the corresponding WIMP-nucleon cross section exceeds the
spin-independent one by several orders of magnitude, what in
general is not the case for the 4th neutrino. The value
$\xi_{loc}$ for Heavy neutrinos is deduced from the result of DAMA/NaI
in term of $\xi_{loc}\sigma_{SI}$, where $\sigma_{SI}$ is the
effective spin-independent WIMP-nucleon cross-section
\begin{equation}
\sigma_{SI}=\frac{G_F^2\mu^2}{8\pi}\frac{\beta_{Na}+\beta_{I}}{V_{Na}^{-2}\beta_{Na}+V_{I}^{-2}\beta_{I}}.
\end{equation}
Here $G_F$ is the Fermi constant, $\mu=mm_{nucl}/(m+m_{nucl})$ is the
reduced mass of neutrino $m$ and nucleon ($m_{nucl}=0.94$ GeV),
$\beta_i=4mm_i/(m+m_i)^2$, $V_i=1-(2-4sin^2\theta_W)Z_i/A_i$ with
$Z_i$ and $A_i$ being numbers of protons and nucleons in nucleus
respectively, $\theta_W$ is the Weinberg angle.

Figure 1 shows a favorable region
for Heavy neutrinos of the 4th generation measured by DAMA and the
fraction corresponding to a Heavy neutrinos contribution to a
local galactic density. The result of DAMA/NaI takes into account
existing uncertainties in DM distribution parameters,  in
form-factor of nuclei, and in the other experimental
parameters \cite{DAMA-review}. The fraction corresponding to a
Heavy neutrinos contribution was estimated by taking
$\Omega_{CDM}=0.3$.
\begin{figure}
\begin{center}
\centerline{\epsfxsize=7cm\epsfbox{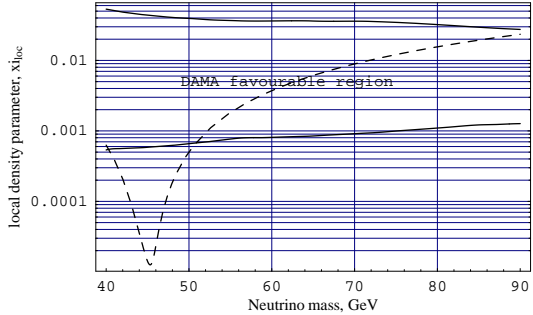}} \caption{Plot of
DAMA favorable region (between upper and lower solid lines) for
Heavy neutrinos of the 4th generation. A dashed line shows the
fraction corresponding to a contribution of the Heavy neutrinos to
CDM of the Universe.} \label{DAMA}
\end{center}
\end{figure}

Note that the present work is based on the essentially updated
DAMA/NaI results in comparison with the previous works \cite{Fargion},
\cite{BK}.

\section{Shadows of Heavy neutrino annihilations in cosmic rays}

Due to a concentration in Galaxy Heavy neutrinos
can annihilate. The products of such annihilation contribute in cosmic
ray fluxes and cosmic gamma radiation. Observational data on cosmic
positrons, antiprotons
and gamma-radiation are sensitive to this contribution.

An analysis of cosmic rays is complicated because a description of
CR production and propagation contains significant uncertainties.
Observational data do not allow to choose parameters of physical
models of CR in a unique way for CR origin (injection spectra of
each CR species) and CR propagation (diffusion coefficients and
their energy dependence, parameters of convection,
re-acceleration, magnetic halo parameters, matter distribution in
Galaxy, model of solar modulation etc.). Recently a detailed study
of CR models was performed in \cite{SM-0},\cite{SM-1},\cite{SM-2}.

In order to study effects of possible DM annihilation
in Galaxy, it is reasonable to accept the most
conservative CR model. Eligible models should reproduce possible
CR data which are the least sensitive to effects of WIMP annihilation
(data on nuclear component of CR, its isotope composition).
"Conventional model (C)" in \cite{SM-1} and "diffusion
re-acceleration model (DR)" in \cite{SM-2} are the most suitable.
We will use in our consideration fluxes of secondary positrons,
antiprotons and gamma-radiation predicted in these models as a
"background". Dark matter annihilation sources will be included
in these models to reproduce DM effects.

An uncertainty in our analysis comes also from unknown
distribution of subdominant Heavy neutrino DM in the Galaxy. There
are many models of distribution of CDM in Galaxy. We will use
models for dominant DM, re-scaling a dark matter density in an
appropriate way for a non-dominant Heavy neutrino DM component .
By fixing the density distribution of DM component in Galaxy we
relate the result of DAMA, sensitive to local density of DM, with
results of CR analysis, sensitive to DM density distribution in
Solar neighborhood.

As a basic model for our estimations we select Evan's halo model,
which in \cite{DAMA-review} was named as C2. The values of
parameters in this model are $v_0=170$ km/sec, $\rho_{loc}=0.67$
GeV/cm$^3$. Density distribution of DM in Galaxy is given by
Eq(41) of \cite{DAMA-review} or Eq(34) of \cite{Belli} with the
parameters $q=1/\sqrt{2}$ and $R_c=5$ kpc
\begin{equation}
\rho(R, z)=const\frac{2R_c^2+R^2}{(R_c^2+R^2+2z^2)^2},
\end{equation}
where $R$ and $z$ are the radius in the galactic plane and
the cylindrical coordinate axis perpendicular to it, $const$ is defined
from a
condition $\rho(R=R_0=8.5 {\rm kpc},z=0)=\rho_{loc}$.

It is worth to note the chosen DM distribution is smooth and does
not have a sharp profile near the Galactic center (GC).

Also, to illustrate a dependence on a halo model choice, we will
consider an isothermal halo model with a sharp behavior of density
near GC. Density distribution of the isothermal halo is given by
\begin{equation}
\rho(R)=\rho_{loc}\frac{R_c^2+R_0^2}{R_c^2+R^2}
\end{equation}
with $R_c=1$ kpc, where $R$ is the distance from GC.

Note that the use of cuspy halo model, like Navarro-Frank-White
model \cite{Navarro}, leads to intermediate results.

\subsection{Signature of Heavy neutrinos annihilation in cosmic gamma
fluxes}

An excess of cosmic gamma-radiation observed by EGRET over
predicted galactic $\gamma$-emission, often called "extragalactic"
$\gamma$-background, can be considered as a possible effect of
dark matter sources. A flux of cosmic gamma-radiation near the
Earth from annihilation of relic Heavy neutrinos is defined by
\begin{equation}
I=\frac{dN_{\gamma}}{dtdSd\Omega dE}=\frac{1}{4\pi}\frac{1}{4}<\sigma
v>\int_0^{\infty}n^2dl\,\frac{dN_{\gamma E}}{dE}.
\end{equation}
Here $<\sigma v>$ is the product of cross section of $N\bar{N}$
annihilation and a relative velocity of neutrinos averaged over
velocity distribution, $n=\rho/m$ is the number density of
neutrinos in Galaxy. An integration is performed along the line of
sight with formally infinite upper limit, $dN_{\gamma E}/dE$ is the
mean multiplicity of photons created in an act of annihilation for
$E-(E+dE)$ energy interval. The factor $1/4$ comes from the fact
that the number densities of neutrinos and antineutrinos are equal
to a half of their total number density $n$. To obtain a
distribution $dN_{\gamma E}/dE$ a code PYTHIA 6.2 was used.

Figures 2 and 3 show $\gamma$-fluxes at the Earth from two
directions: from the Galactic center (Fig.2) and from a halo in
the direction of Galactic zenith (Fig.3). Corrected EGRET data for
the halo were taken from \cite{SM-3}, where EGRET data were
re-analyzed in an advanced approach in which the galactic
contribution was subtracted giving pure isotropic "extragalactic"
$\gamma$-radiation. Dashed/solid lines in these figures show
annihilation/annihilation plus background $\gamma$-fluxes for
neutrino mass ranging 47-80 GeV. The annihilation
$\gamma$-fluxes were obtained selecting density parameter
$\xi_{loc}$ at given (Evan's) density distribution to fit in the
best way the observation data for the accepted value of neutrino
mass. In a fitting procedure $\chi^2$ criterion was used by fixing
the Galactic contribution (background) and changing the
annihilation flux by the $\xi_{loc}$-parameter. Neutrino masses
were chosen as 47, 50, 55, 60, 65, 70, 75, 80 GeV.

The low energy part of $\gamma$-spectrum from halo (Fig.3) is not
reproduced by annihilation of Heavy neutrinos in the halo. However
this low energy part can be explained by extragalactic emission
based on a blazar population \cite{SM-3},\cite{neutralino-gamma}.
This emission is expected with a similar slope in spectra as data
points exhibit. Extraction of pure "extragalactic" $\gamma$-flux
in direction of Galactic center (GC) is more complicated problem.
We used observational data and a prediction of galactic
contribution for $\gamma$-flux from GC \cite{SM-1} in accordance
with model "C". The galactic contribution (refereed to as a
background) is shown in Fig.2 by dot-dashed line.
\begin{figure}
\begin{center}
\centerline{\epsfxsize=7.5cm\epsfbox{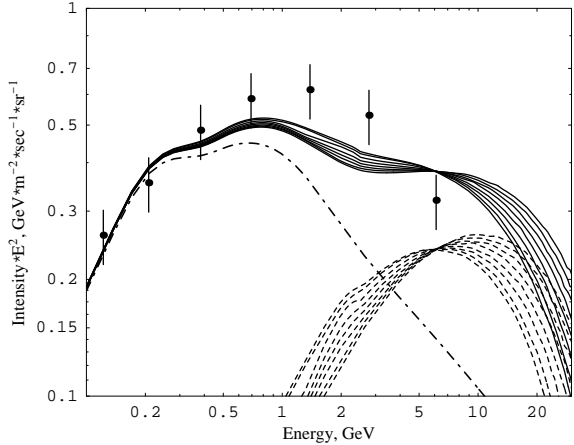}}
\caption{Cosmic gamma-radiation from galactic center
($0.5^o<l<30.0^o,330.0^o<l<359.0^o$): EGRET data, predicted
background (dot-dashed line), and the best-fit contribution from
47-80 GeV neutrino DM for Evan's halo model. The set of dashed
lines corresponds to pure annihilation gamma-fluxes, the set of
solid lines is the sum of background and annihilation
fluxes.\label{fig.2}}
\end{center}
\end{figure}
\begin{figure}
\begin{center}
\centerline{\epsfxsize=7.5cm\epsfbox{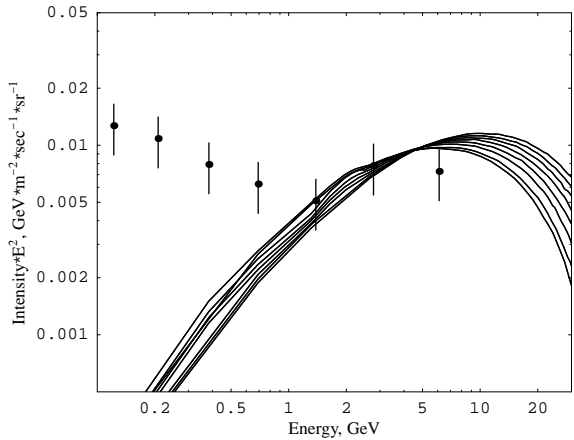}}
\caption{Cosmic gamma-radiation from zenith galactic direction:
EGRET data and the best-fit gamma-flux from 47-80 GeV neutrino
DM annihilation (the set of lines) for Evan's halo model.
\label{fig.3}}
\end{center}
\end{figure}

All $\xi_{loc}$-parameters, fitting in the best way the predicted
annihilation with background fluxes to observational data, will be
presented in a figure below.

\subsection{Signature of Heavy neutrinos annihilation in cosmic $e^+$ and
$\overline{p}$ fluxes}

The flux of charged particles from annihilation of Heavy neutrinos
in Galaxy near the Earth is defined by a diffusion of particles
from $N\bar{N}$ annihilation to the region around the Earth of the
size as large as the characteristic diffusion length. For antiprotons,
which do not experience significant energy loss, this region is
determined by the size of halo where they are trapped by magnetic
field. For positrons the size of region of dark matter
annihilation sources contributing to a flux near the Earth
depends on energy loss of positrons. This does not allow
positrons, created with an energy $E_0$, to come with energy $E$
from distance strongly exceeding
\begin{equation}
\lambda(E,E_0)=\left(\int_{E_0}^E\frac{D(E')}{b(E')}dE'\right)^{1/2}.
\end{equation}
Here $D$ is the diffusion coefficient which is energy (rigidity)
dependent, $b(E)$ is the rate of energy loss defined as
\begin{equation}
\frac{dE}{dt}=-b(E).
\end{equation}

Charged cosmic particles experience a solar modulation. To take
into account this effect we will use force-field model
\cite{solar modulation}. In this approximation the intensity
measured at the top of the Earth's atmosphere (inner heliosphere) at
the energy $E_{Earth}$ corresponds to a local interstellar (LIS)
intensity (outer heliosphere) through the relation
\begin{equation}
I_{LIS}(E=E_{Earth}+\Phi(t))=\frac{(E_{Earth}+\Phi(t))^2-m_p^2}{E_{Earth}^2-m_p^2}
\,I_{Earth}(E_{Earth},t).
\end{equation}
Here $m_p$ is the mass of cosmic particle, $\Phi(t)$ is the
energy lost by the cosmic particles during their travel in the
heliosphere. $\Phi(t)$ is the parameter of the model which can be
derived from an observation of appropriate period of Solar
activity. A dependence of solar modulation of CR on the sign of
particle charge appears at low energy (at LIS
energy, less than 1-2 GeV). It is shown in \cite{Casadei} this dependence
during
nineties (positive half-cycle of the Sun) broke the force-field
approximation for description of data on negatively charged
particles (electrons, antiprotons) in low energy range. Whereas
positively charged particles were well described by the
force-field model in that period. We will use data on cosmic
positrons and antiprotons transformed into LIS by Eq(8), what
will lead to some underestimation of LIS antiproton flux at low
energy.
Note that the energy scale of modulation is detemnined by $\Phi(t)$,
being around $1$GeV, and for $E>>1$ GeV effect of
modulation is negligible.

For an estimation of positron flux from $N\bar{N}$ annihilation
we adopted the diffusion approximation of positron propagation in
Galaxy without inclusion of the effect of diffusion zone
boundaries \cite{Ginzburg}. It is well-known that a strong energy
loss of high energy cosmic $e^{\pm}$ makes their spectra
dependent on the space distribution of density of
$e^{\pm}$-sources. It disfavors the use of "leaky-box" model for
a quantitative estimation of the effects of tangled, diffusion
propagation of $e^{\pm}$ in Galaxy. A diffusion coefficient and
energy loss parameter were chosen following "DR" model \cite{SM-2}
\begin{eqnarray}
D(E)=6.1\times10^{28}\left(\frac{E}{4\,{\rm GeV}}\right)^{0.33}\,{\rm cm^2s^{-1}},\\
b(E)=\beta E^2,\,\,\,\beta=1.52\times10^{-9}(0.5+0.5(H/3\,{\rm \mu G})^2)\,{\rm yr^{-1}GeV^{-1}},
\end{eqnarray}
Here in the expression for $\beta$ the dependence on the averaged galactic magnetic
field $H$ and its value $3\,{\rm \mu G}$ are taken in accordance with \cite{Turner}.
Such parameters (Eq(9-10)) allow positrons originated as far as
at GC to contribute to the flux near the Earth. A background
(secondary) positron flux was taken as predicted in "DR" model
keeping accordance with the choice of parameters (Eq(9-10)).
"DR" model takes into account effect of boundaries and effect of
re-acceleration, acceleration of cosmic particles (initially
accelerated in their sources) during their propagation in
interstellar medium. A disagreement between our estimation of
annihilation flux and the used prediction of background is not
significant. An effect of boundaries of diffusion zone is not
very important for Evan's DM halo model as it is seen from
\cite{SM-4}.

The decrease of halo size $z_h$ leads to the de-population of
distribution of positrons with $\lambda>z_h$ (Eq(6)) because they
escape the diffusion zone more intensively than at larger $z_h$
and also because a contribution from dark matter annihilation
sources situated outside the diffusion zone is not taken into
account. The escape from the diffusion zone leads to diminishing
relative contribution of annihilation positrons from GC with a
decreasing halo size, what is more marked effect for halo models
with a sharp density profile near GC. The use of Evan's halo model
with its smoothed density distribution provides better accuracy
for the used approximation than other models with sharper
profiles. Also to reduce a deviation of annihilation fluxes
obtained in our approximation from that one which would be
predicted in "DR" model, induced by the neglecting of the
boundaries, we excluded a contribution from dark matter
annihilation sources situated outside the diffusion zone of
$z_h=4$ kpc. Effect of re-acceleration appears below 5 GeV in
positron spectra \cite{SM-4} (or the right part of Fig.5 in the
ref. \cite{reacceleration}), where the role of secondary positrons
shades the possible dark matter annihilation sources contribution.

In Fig.4 the predicted LIS positron fluxes as compared to
observational data of HEAT \cite{HEAT} are presented. There are
the secondary positron flux (dot-dashed line), the sum of
secondary and the best-fit annihilation fluxes (the set of solid
lines) and separately the last ones (the set of dashed lines),
%{\bf We do not need c in formulas below}
obtained for the range of neutrino mass 47-80 GeV. Note, that
the curves corresponding to the annihilation fluxes are extended
up to the energy equal to $m$ of neutrino, above this energy
the predicted total fluxes (secondary plus annihilation positrons)
are the same as the secondary flux at $E>m$.
%{\bf A sharp "step" in form of spectra would have to be washed
%a little due to re-acceleration effect.}
HEAT data were
"demodulated" taking $\Phi=664$ MeV, derived from data of CAPRICE
\cite{Casadei}. A fitting was performed in the same way as
described above for gamma-radiation. First point of HEAT (slightly
below 2 GeV in Fig.4) was omitted in the procedure of fitting.
This point is apparently inconsistent even with the predicted
secondary positron flux and also inconsistent with the other
measurements of cosmic positrons \cite{Casadei}. Note that the low
energy range should be considered cautiously, because predictions
in this range depend on possible effects of re-acceleration.

Let's consider an effect of diffusion coefficient variation, or
correspondingly (see Eq(6)) a variation of positron diffusion
length. A decrease in $D$ by a factor 10 modifies a little a slope
(makes steeper) of spectra of annihilation positrons near the
Earth and changes the best-fit $\xi_{loc}$ parameter by less than 20\%.
The further decrease in $D$ causes no influence, because mean free path
length $\lambda$ becomes less than typical physical scales of the
problem (the distance to the nearest boundary of diffusion zone or
the length scale of variation in the DM density distribution).
\begin{figure}
\begin{center}
\centerline{\epsfxsize=7.5cm\epsfbox{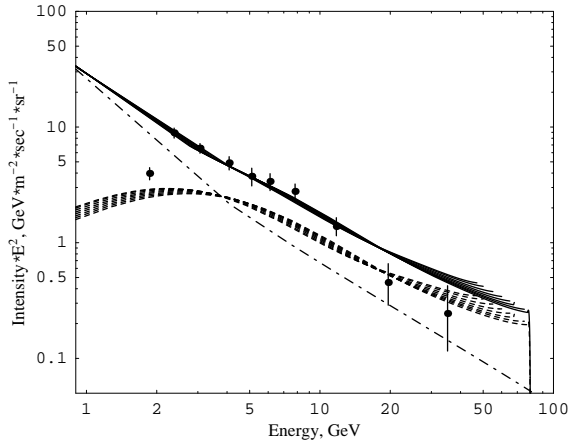}}
\caption{Cosmic positrons (LIS): HEAT data, predicted background
(dot-dashed line), and the best-fit contribution from 47-80 GeV
neutrino DM (the set of dashed lines is pure annihilation positron
fluxes, the set of solid lines is the sum of background and
annihilation fluxes) for Evan's halo model. \label{fig.4}}
\end{center}
\end{figure}

To estimate  the flux of antiprotons from  $N\bar{N}$ annihilation
near the Earth we accept a leaky-box model. In this model the
intensity can be defined by a simple expression
\begin{equation}
I=\frac{v_{\bar{p}}}{4\pi}<\dot{n}_{\bar{p}}>\tau_{conf}.
\end{equation}
Here $v_{\bar{p}}$ is the velocity of antiproton,
$<\dot{n}_{\bar{p}}>=\frac{\int\frac{1}{4}n^2<\sigma
v>dV}{V_{Gal}}\frac{dm_{\bar{p}}}{dE}$ is the mean number of
antiprotons created in a unit volume per second per energy
interval $E-(E+dE)$, averaged over the volume of Galaxy,
$\tau_{conf}$ is the confinement time, the other notations are
analogous to those introduced in Eq(5). Time $\tau_{conf}$ is a
parameter of the model and it has the meaning of the time of
$\bar{p}$ confinement in Galaxy. We chose $\tau_{conf}$ to be
$10^7$ years as in the early works \cite{Fargion}. The volume of
Galaxy is supposed to be the volume of region where antiprotons
are confined and this region is chosen in form of a disk with
radius 25 kpc and semiheight $z_h=4$ kpc, typical for CR models.
Energy losses of antiprotons are neglected in Eq(11). Such
simplification is justified by small mean matter column
(5 g/cm$^2$) traveled
by cosmic nuclei, , which was deduced from CR analysis.
A small fraction of antiprotons, which lose their energy in an
inelastic scattering on protons of medium ("tertiary" component),
appears in antiproton spectra at very low energy \cite{SM-2}.

As a background the secondary antiprotons predicted in model "C" of
\cite{SM-1} were used. For a comparison with observations the
combined data of BESS'95 and '97 were used \cite{BESS}, which were
demodulated with parameter $\Phi=540$ MeV, derived from BESS'95
\cite{BESS-all}. Data BESS'98 belong to time of high solar
activity what is less suitable from point of view of detection of
possible dark matter annihilation sources \cite{BESS-all}. Figure
5 shows a spectrum of cosmic antiprotons. Unlike the cases of
gamma-radiation and positrons the spectra of antiprotons from
$N\bar{N}$ annihilation for different values of Heavy neutrino
mass coincide within an energy interval presented in Fig.5. A
charge-sign dependence of solar modulation and an effect of
re-acceleration are significant at low energy, making results in
this energy range less certain.
\begin{figure}
\begin{center}
\centerline{\epsfxsize=7.5cm\epsfbox{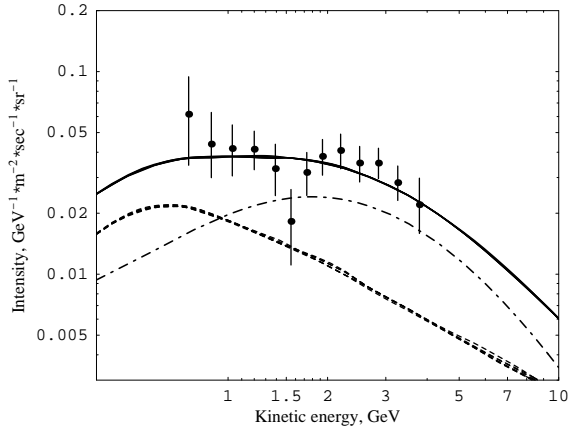}}
\caption{Cosmic antiprotons (LIS): BESS(95+97) data, predicted
background (dot-dashed line), and the best-fit contribution from
47-80 GeV neutrino DM (the set of dashed lines is pure
annihilation antiproton fluxes, the set of solid lines is the sum
of background and annihilation fluxes) for Evan's halo model.
Note, that for the considered interval of neutrino masses
the sets of dashed and solid lines are virtually reduced to
single lines.
\label{fig.5}}
\end{center}
\end{figure}
%
%
%{\bf I see single lines not a set of lines. May be it is better to say
%something.}

An increase of parameter $\tau_{conf}$ or/and a decrease of volume
comprising $\bar{p}$ propagation zone in Galaxy lead to a decrease
in the best-fit density parameter $\xi_{loc}$. Uncertainties in
$\tau_{conf}$ and $V_{Gal}$ lead to overall uncertainty of about a
factor 2.

Note, that the analysis of CR carried out in this work differs
from analogous analysis performed in the previous works
\cite{Fargion}, \cite{4thN-CR} by more refined consideration, in
particular more realistic CR models and models of distribution of
Heavy neutrinos in Galaxy.

\section{Heavy neutrino in underground versus cosmic rays signals}

As one can see from figures 2-5, the presence of dark matter
annihilation sources in the form of Heavy neutrinos improves
description of existing data on cosmic gamma-radiation, positrons,
antiprotons with corresponding $\xi_{loc}$ selected in the best
way from observational data. The annihilation fluxes in these
figures were obtained for Evan's halo model (Eq(3)) as described
above. In the same manner as in case of Evan's model the
parameters $\xi_{loc}$, allowing to fit in the best way CR data,
were obtained for isothermal halo model (Eq(4)). All these
parameters for different values of neutrino mass are shown in
Fig.6 in comparison with those preferable in measurements of DAMA.
There is the set of black lines in upper half of figure, starting
at $m=46$ GeV and ending at $m=80$ GeV, which corresponds to
$\xi_{loc}$ parameters, inferred from CR analysis. Pairs of solid
(dot-dashed), dotted and dashed lines of this set are related with
the best-fit $\xi_{loc}$ for gamma-radiation from halo (GC), for
cosmic positrons and antiprotons respectively. Upper and lower
lines of each pair correspond to Evan's and isothermal halo models
respectively. Solid and dashed grey lines, going across the
picture, enclose favorable region of DAMA. For consistent
comparison of $\xi_{loc}$, inferred from CR analysis using Evan's
halo model, the values of $\xi_{loc}$, derived from analysis of
DAMA/NaI measurements based on the same Evan halo model, are shown
by dashed grey lines.

%{\bf I change caption a little bit}
%
\begin{figure}
\begin{center}
\centerline{\epsfxsize=8cm\epsfbox{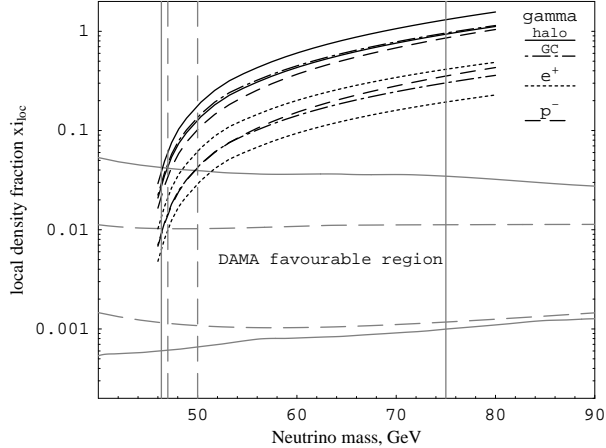}}
\caption{DAMA favorable region (as in Fig.1) and
the best-fit density parameters deduced from cosmic
gamma-radiation (from halo and CG), positron and antiproton
analysis. Horizontal grey dashed and solid lines enclose DAMA
favorable region accepting Evan's halo model and other halo models,
respectively. The set of upper lines corresponds to the $\xi_{loc}$
parameters preferable for CR data. In this set of lines, upper and
lower lines of the same type correspond to Evan's halo model and
to isothermal halo model, respectively. Vertical grey dashed and
solid lines restrict $1\sigma$ and $2\sigma$ allowable range of
the 4th neutrino mass deduced from the particle physics data analysis.
\label{fig.6}}
\end{center}
\end{figure}
%
%{\bf ``many'' is not good}

All $\xi_{loc}$ parameters, obtained from CR analysis, define
parameters favored by CR data as well as upper constraints
imposed by CR data. So, given results allow to make a conclusion
that CR data are consistent with measurements of DAMA/NaI treated in
framework of hypothesis about the 4th generation neutrino.

An additional
source of information about possible existence of WIMPs is data
from the search for light neutrino fluxes from annihilation of
WIMPs accumulated inside the Earth and Sun. But existing data of
underground measurements of neutrino fluxes exhibit, contrary to
CR data, a lack of neutrinos as compared to predicted background
(atmospheric neutrinos). New physics is possibly required here. An
interpretation based on 3-flavor neutrino oscillations fails to
reproduce all appropriate existing data without an
introduction of a new sterile
neutrino. An estimation of muon neutrino fluxes from annihilation
of Heavy neutrinos inside the Earth gives a result comparable with
the expected corresponding atmospheric neutrino flux in the energy
range $>3$ GeV for $m=50$ GeV for acceptable parameters of Evan's
halo model. In this analysis the result depends also on WIMP
velocity distribution which affects the capture rate of WIMPs by
the Earth. This ratio is reduced by a few times if the velocity
distribution given by Evan's halo model (Eq(A1-A4) \cite{Belli})
with velocity parameter $v_0=170$ km/sec is replaced by Maxwellian
distribution with r.m.s. velocity $v_0=220$ km/sec. An analysis of
underground measurements of upward-going muons (Super-Kamiokande,
MACRO, Baksan), induced by neutrino fluxes, requires its further
development.

A question about an agreement between all predicted parameters for
the 4th generation neutrino, helping to improve description data
of different species, is of greater interest. Striking is a
relative closeness of $\xi_{loc}$ parameters preferable for
different observations.
%{\bf [all uncertainties affecting given results need to be minded -?]}.
Figure 6 shows that "play" with the form of density distribution
of Heavy neutrinos in Galaxy is able to change significantly
$\xi_{loc}$ derived from different observations. An agreement
between the considered data is possible. For the chosen isothermal
model all $\xi_{loc}$ parameters (lower lines from pairs of lines
of upper (black) lines in Fig.6) favored by CR data, a neutrino
mass is close to its lower constraint, in the region of
corresponding magnitudes, deduced from DAMA experiment accounting
for different halo models.

%{\bf I do not understand the previous sentence!}
In the case of isothermal model, the
value of $\xi_{loc}$ parameter inferred from analysis of cosmic
gamma-radiation from halo (solid line), differs from the other
ones. This discrepancy can be due to a possible extragalactic
$\gamma$-radiation (see Fig.3) the account for which can lead to
better agreement. But, of course, results of indirect WIMP
searches should be treated cautiously taking into account the low
precision of the corresponding experimental data.

Given Evan's halo model, predictions of $\xi_{loc}$ from the data
on different CR species are differed by a factor of three. In a
view of approximations in CR analysis described above it should
not be considered as the principal discrepancy. An agreement achieved
between predictions of parameters preferable for CR data and for
measurement of DAMA would require reduction of parameter
$\xi_{loc}$, preferred by CR data, by a factor a few - ten %{\bf(???)}
in the allowed range of neutrino mass below 50 GeV. Such
a reduction corresponds to an amplification of the annihilation
flux proportional to the square of that factor.

In other words, the results of DAMA/NaI experiment are compatible with
indirect effects of 4th neutrino annihilation, but the observational
indications to WIMP annihilation effects in cosmic rays and gamma
radiation can be explained together with these results only for some models
and for a very narrow interval of neutrino masses (46-47 GeV). To increase
the range of neutrino masses, at which direct and indirect WIMP signals
can find simultaneous explanation,
the rate of 4th neutrino annihilation in Galaxy should be much larger.

\section{Three ways to extend neutrino models and mass range}

\subsection {Amplification of neutrino annihilation due to
clumpiness}

There is a possibility to amplify CR flux created by dark matter
annihilation sources maintaining local density and average density
distribution of DM in Galaxy. This possibility is related with a
clumpy DM distribution in Galaxy (\cite{Berezinsky} %(add Berquist)
and references therein). In particular we are considering local
clustering in our galactic center and halo with no peculiar
density enhancement for our neighborhood. The opposite situation
(higher Solar and lower global galactic density) is a
possibility much less probable and attractive.

CDM might form clumps on the stage of structure formation in the
Universe. As it was shown in \cite{Berezinsky} a small fraction of
total DM mass (a few$\times10^{-3}$) can survive to present time
in form of clumps and it is enough to provide strong enhancement
(up to a few orders of magnitude) of annihilation signal. Being
a non-dominating DM component, Heavy neutrinos most likely
do not form their own clumps. A formation of clumps should be governed
by the dominating CDM component. Heavy neutrinos should subserve in
such processes for values of the formed clump mass, $M_{clump}$,
exceeding some minimal one, $M_{N\,min}$. The last one is defined by
the size of a proto-clump equal to free-streaming length of
neutrinos, $\lambda_{fs}$, when inhomogeneities start to grow.
$\lambda_{fs}$ depends on the moment of  Heavy neutrinos
decoupling from an ambient plasma. For 50 GeV neutrino the
temperature of decoupling is estimated as $T_d\sim
20$ MeV, %\cite{Khlopov},
%This temperature was calculated a lot of time ago. I do not
%understand this reference
whereas the mass $M_{N\,min}\sim 0.6\times10^{-6}M_{\odot}$ (Eq(37)
of \cite{Berezinsky}). Provide a dominant CDM component has a less
free-streaming scale (that is, for instance, quite probable for
neutralinos and heavy gravitinos), there would exist the clumps with
masses in the range $M_{clump}>M_{min}$ so that
$M_{min}<M_{N\,min}$ (for neutralino $M_{min}$ is estimated as
$\sim10^{-8}M_{\odot}$ \cite{Berezinsky}).
For Heavy neutrinos the
ranges of clump masses are $M_{min}<M_{clump}<M_{N\,min}$ and
$M_{clump}>M_{N\,min}$. Creation of clumps only of the second mass
range is not expected to proceed with a separation of the dominant CDM
component and Heavy neutrinos. We will suppose a conservation of
proportionality (ratio) between densities of dominant CDM
component and Heavy neutrinos in a such clump creation. Clumps
lighter than $M_{N\,min}$, if they are, can be populated by Heavy
neutrinos in less degree in accordance with the mechanism of
adiabatic loss of energy by collisionless particles (neutrinos) in
an external variable gravitation field \cite{ZKKC}.

Estimations of \cite{Berezinsky} for the enhancement factor of
dominant DM annihilation flux due to the presence of clumps can be
applied to Heavy neutrinos. This factor is defined as
\begin{equation}
\eta=\frac{I_{clump}+I_{hom}}{I_{hom}}
\end{equation}
where $I_{clump}$ and $I_{hom}$ are the intensities of
annihilation fluxes from clumps of DM and homogeneously distributed
DM. It crucially depends on minimal mass of clumps,
$\eta(M_{min})$, the lightest clumps give the main contribution
into an annihilation rate. Under assumption on proportionality
between densities the predictions for enhancement factor can be
referred to non-dominating Heavy neutrinos DM assuming the minimal
clump mass to be $M_{N\,min}$
\begin{equation}
\eta_N=\eta(M_{N\,min}).
\end{equation}
Note, that Fig.1 implies some deviation from such proportionality
within a factor 1-100 at $m=50$ GeV for an average local density
(which is deduced from dynamical observation to be within 0.17-1.7
GeV/cm$^3$), if real local density (in vicinity of the Earth) does
not differ significantly from the average one.

Additional contribution into an enhancement of the neutrino
annihilation rate will result from the clumps with a mass in the
range $M_{min}-M_{N\,min}$. The enhancement from these clumps can
be roughly estimated as a corresponding estimation in
\cite{Berezinsky} for $M_{min}$ reduced with respect to smaller
relative compression of Heavy neutrino density inside the given
clumps as compared to that of dominant DM. In each such clump the
relative compression of neutrino density, i.e. the ratio of
densities inside a clump and outside it (of homogeneous component
near the clump), should be smaller than that of dominant component
of matter (CDM) in accordance with \cite{ZKKC}
\begin{equation}
\frac{\rho_{N\,clump}}{\rho_{N\,hom}}=\left(\frac{\rho_{CDM\,clump}}{\rho_{CDM\,hom}}\right)^{3/4}.
\end{equation}
As a first approximation we neglect details (differences) of
density distributions of dominant component and Heavy neutrinos
inside these clumps. So, the factor $\eta$ is determined by
squared ratio of densities above both for Heavy neutrinos and for
CDM. The enhancement factor for Heavy neutrinos for a clump mass
between $M_{min}$ and $M_{N\,min}$ can be estimated as
\begin{equation}
\eta_{N\,add}=\eta(M_{min})^{3/4},
\end{equation}
where $\eta(M_{min})$ is the enhancement factor as predicted in
\cite{Berezinsky} (for dominant DM component).

Factor $\eta$ increases with a decrease of clump mass
\cite{Berezinsky}, so a contribution from the clumps of mass
$M_{min}<M_{clump}<M_{N\,min}$, which are relatively less
populated by Heavy neutrinos, can be comparable with those of mass
$M_{clump}>M_{N\,min}$. For instance, for an index of primeval
perturbation spectrum $n_p=1.05$, index of power-like density
distribution inside the clump $\beta=1.8$ at
$M_{N\,min}=0.6\times10^{-6}M_{\odot}$ we have $\eta_N\approx20$
(see Fig.5 of \cite{Berezinsky}), and assuming
$M_{min}=10^{-8}M_{\odot}$ we obtain
$\eta_{N\,add}\approx(30)^{3/4}\approx13$. For essentially smaller
values of $M_{min}$ the contribution from clumps with
$M_{min}<M_{clump}<M_{N\,min}$ is turned out to prevail.

A density parameter $\xi_{loc}$ should decrease as a square root
of enhancement factor $\eta_N$, so in example above we get
$\xi_{loc}$ by a factor 5 less at 50 GeV than it is in Fig.6. For
other values of neutrino mass of interest the result is virtually
the same. An agreement between measurement of DAMA/NaI and CR
observation data is possible for the hypothesis of 4th neutrino
with Evan's halo model within allowed neutrino mass range below 50
GeV (see Fig.7).
%
%{\bf I change caption a little bit}
%
\begin{figure}
\begin{center}
\centerline{\epsfxsize=8cm\epsfbox{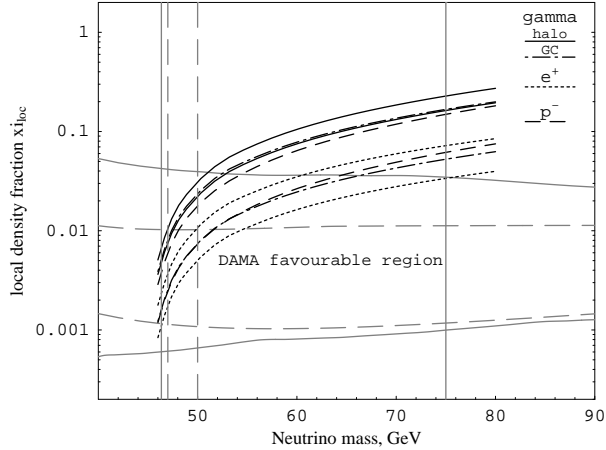}}
\caption{Effect of neutrino clumpiness. All notations are analogous to those
of fig. 6.
%DAMA/NaI favorable region (as in Fig.1) and
%the best-fit density parameters deduced from cosmic
%gamma-radiation (from halo and CG), positron and antiproton
%analysis. Horizontal grey dashed and solid lines enclose DAMA/NaI
%favorable region accepting Evan's halo model and other halo models,
%respectively. The set of upper lines corresponds to the $\xi_{loc}$
%parameters preferable for CR data. In this set of lines, upper and
%lower lines of the same type correspond to Evan's halo model and
%to isothermal halo model, respectively. Vertical grey dashed and
%solid lines restrict $1\sigma$ and $2\sigma$ allowable range of
%the 4th neutrino mass deduced from particle physics data analysis.
\label{fig.7}}
\end{center}
\end{figure}

Note that in our approximation we did not pay attention to a
correction for given quantitative estimations of an effect of
clumps due to a different DM density distribution in Galaxy
(Evan's) than it was supposed in \cite{Berezinsky}. Also note,
that a destruction of clumps near GC \cite{Berezinsky} should
partially decrease the enhancement of annihilation $\gamma$-flux
from GC due to clumps. This may improve agreement between results
for photons from GC and the halo.

In the case if a minimal possible DM clump mass, $M_{min}$, is
greater than $M_{N\,min}$ (there is only a unique mass range),
then the enhancement factor would be given by Eq(13) with
$M_{min}$ instead of $M_{N\,min}$.

\subsection {Amplification of neutrino annihilation due to
new Coulomb-like interaction}

An annihilation signal as a signature of existence of DM particles
like Heavy neutrinos implies a condition of the presence of both
particles and antiparticles. The case considered in the present
article was based on
an assumption of charge symmetry of 4th generation particles,
i.e. an equality between the numbers of primordial 4th neutrinos and
antineutrinos. Such a statement can find physical foundation in
superstring models. New charge(s) is(are) predicted there which,
being strictly conserved, can be ascribed only to 4th generation
particles \cite{Shibaev}. It accounts for absolute stability of
the lightest particle bearing this charge (assumed to be the 4th
neutrino) and an equality between particles and antiparticles of a
new generation.

An important consequence of a new charge is an effect of new
interaction. In a wide class of models this charge is $U(1)$-gauge
charge which leads to existence of corresponding massless gauge bosons
($y$-photons) and to a Coulomb-like interaction of 4th neutrinos. It was
revealed in \cite{Sakharov
enhancement}, that this new
interaction does not influence significantly
the results of 4th neutrino freezing out in early Universe, but it can
increase their annihilation
signal in Galaxy by a few hundred
times. The matter is that in the considered range of neutrino masses
ordinary electroweak
$Z$-bozon resonance annihilation channel dominates over the new channel of
2$y$annihilation
and the main effect of new interaction is Coulomb-like factor in the cross
sections
of slow charged particles. Such factor, $C$, first deduced by A.Sakharov
\cite{Sakharov} for
Coulomb interaction of slow electrically charged particles, in the case
of Coulomb-like interaction with "fine structure constant" $\alpha_y \sim
1/137 - 1/14$
has the form \cite{Sakharov enhancement}
\begin{equation}
C=\frac{2 \pi \alpha_y c/v}{1 - \exp({-2 \pi \alpha_y c/v})},
\end{equation}
where $c$ is the speed of light and $v$ is the relative velocity of charged
particles.
At $v/c \ge 1/10$, what is the case for the period of 4th neutrino freezing
out in the early Unvierse,
this factor is close to 1, but for $v/c \ll 1$, being the case for
neutrinos in Galaxy, it increases
their annihilation rate by the factor of Sakharov's enhancement
\cite{Sakharov enhancement}
\begin{equation}
C(v)={2 \pi \alpha_y \frac{c}{v}} \simeq 10^2
\left(\frac{\alpha_y}{1/60}\right)\left(\frac{300\,{\rm km/s}}{v}\right).
\end{equation}

It would provide a decrease of predicted the best-fit
$\xi_{loc}$ parameters as square root of that enhancement
{\bf (being properly averaged over velocity distribution).}
%The account for both clumpiness and for this new  Coulomb-like
%interaction, possessed by 4th neutrinos, extends the possibility
%of unified explanation of CR and DAMA/NaI data in the framework
%of 4th neutrino hypothesis practically for the whole considered
%interval of neutrino masses, excluding values of mass around and
%below 50 GeV for some models (see Fig. 8).
The account for this new  Coulomb-like
interaction, possessed by 4th neutrinos, extends the possibility
of unified explanation of CR and DAMA/NaI data in the framework
of 4th neutrino hypothesis for the most part of considered
interval of neutrino masses. Figure 8 corresponds to the case of
$\alpha_y=1/30$, which is the most natural value for this coupling constant.
\begin{figure}
\begin{center}
\centerline{\epsfxsize=8cm\epsfbox{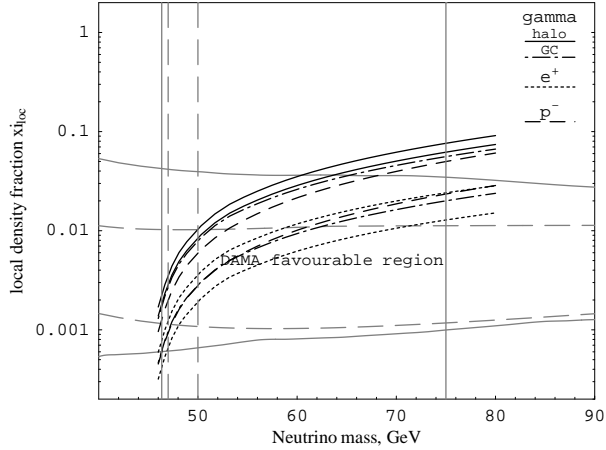}}
\caption{Effect of neutrino Coulomb-like new
interaction. All notations are analogous to those
of fig. 6.
%DAMA/NaI favorable region (as in Fig.1) and
%the best-fit density parameters deduced from cosmic
%gamma-radiation (from halo and CG), positron and antiproton
%analysis. Horizontal grey dashed and solid lines enclose DAMA
%favorable region accepting Evan's halo model and other halo models,
%respectively. The set of upper lines corresponds to the $\xi_{loc}$
%parameters preferable for CR data. In this set of lines, upper and
%lower lines of the same type correspond to Evan's halo model and
%to isothermal halo model, respectively. Vertical grey dashed and
%solid lines restrict $1\sigma$ and $2\sigma$ allowable range of
%the 4th neutrino mass deduced from particle physics data analysis.
\label{fig.8}}
\end{center}
\end{figure}

The case of clumpiness of Heavy neutrinos with new interaction, while
being possible, is more complex and involves both factors ($\eta$ and
$\alpha_y$). Its combined role will offer more tunable scenarios;
however for sake of simplicity we wil not take it into account here.

On the first sight Sakharov's enhancement does not influence the annihilation
rate of
accumulated 4th neutrinos inside the Earth and Sun, which is
defined by their capture rate. However, the existence of
stable quark of 4th generation can make the picture more complicated.
Being compatible \cite{Shibaev} with the constraints on the abundance
of anomalous isotopes, the presence of small amount of anomalous hadrons,
containing this quark (or antiquark) and possessing
the $U(1)$-gauge charge can cause asymmetry in capture rates for 4th
neutrino
and antineutrino and thus influence their annihilation rate
in Earth and Sun.  A role of this gauge interaction,
its charges and field in
effects of Heavy neutrinos as well as the account of possible
existence of stable 4th generation hadrons requires a separate discussion.

\subsection {The role of Heavy neutrino asymmetry and decays}

If neutrinos of 4th generation do not possess new $U(1)$-gauge
charge, there does not appear any fundamental reason for their
absolute stability as well as for their strict charge symmetry.
If charge symmetry is absent, which is the case for
baryons in the Universe, the 4th neutrinos would prevail over
their antineutrinos (or vise versa) and they could decay. A
magnitude of this asymmetry, being defined as the ratio of
difference of present number densities of 4th neutrinos and 4th
antineutrinos ($\delta n$) and present relic photon number
density, is an additional parameter of the problem.\footnote{In the case
of light neutrinos, which decouple in the conditions of thermodynamical
equilibium, such asymmetry would determine their chemical potential
(see review in \cite{Dolgov}).}
This parameter
should be much less than that of baryons in order not to exceed an
essentially relative contribution of 4th neutrino into the density
of CDM derived from measurements of DAMA (Fig.1). A difference
$\delta n\sim {\rm a\, few}\times n_{sym}$, where $n_{sym}$ is the
relic Heavy neutrino density in the case of charge symmetry,
leads to an increase of relic 4th neutrino density by a few times
and to an exponential decrease of density of relic 4th
antineutrinos by a few orders of magnitude. As seen in Fig.1 this
would be especially favored for neutrino mass around 50 GeV. If
relic neutrinos survive to present time, their annihilation
signals from Galaxy and from the Earth and Sun weaken by a few
orders of magnitude as compared to the symmetric case.

However, in the asymmetric case another signature of 4th neutrino DM in CR
is possible.
Neutrinos of 4th generation are unstable in this case
and their decays in the galactic halo can lead to effects similar to the
ones from
stable neutrino annihilation.
In this case CR data can be reproduced,
if 4th neutrino decay lifetime is, for Evan's
density distribution,
%
%\begin{equation}
%\tau=\frac{0.7\times10^{17}{\rm\, years}}{\xi_{loc}}\frac{50\,{\rm
%GeV}}{m}\frac{F_Z(m)}{F_Z(m=50\,{\rm GeV})}.
%\label{unutime}
%\end{equation}
%
\begin{equation}
\tau\approx \xi_{loc}\cdot (0.2-2)\times10^{19}{\rm\, years}\frac{50\,{\rm
GeV}}{m}.
\label{unutime}
\end{equation}
%
%Here $F_Z(m)=\left[4m^2/m_Z^2-1\right]^2+\Gamma_Z^2/m_Z^2$,
%$F_Z(m=50\,{\rm GeV})=0.0418$ with $m_Z$ and $\Gamma_Z$ being the
%mass and decay width of $Z$-boson.
Here $\xi_{loc}$ is the local density fraction of neutrinos in charge
asymmetric case.
The value of the decay rate preferable for CR data changes weakly with
variation of neutrino mass
(since it is defined by the observed CR fluxes), so uncertainty 0.2-2 in
estimation (\ref{unutime}) is mainly induced by uncertainty of best-fit
$\xi_{loc}$
parameters deduced from data of different CR species (for Evan's halo
model).
Note that there is a difference
by a factor 2 in energy release for annihilation reaction and
decay process. But in the numerical estimations for the effects of decay
the predictions for CR fluxes,
induced by neutrino annihilation, can be used without significant change
with only proper account for the change in the energy release,
provided that hadron modes are present in decay with sufficient
probability.
Parameter $\xi_{loc}\approx 0.01 - 0.001$ would provide for neutrino
of mass about 47-70 GeV with the lifetime given above
an agreement between the measurement of DAMA/NaI and CR data.
So, preferable lifetime of 4th neutrino is
$\tau\sim (2\times10^{15}-2\times 10^{17})$ years.
Note, that clumpiness does not affect the
fluxes of products of Heavy neutrino decay in the Galaxy. Also
note, that the annihilation rate of neutrinos accumulated inside
the Earth and Sun in the symmetric case corresponds to a timescale
less than the age of Solar system (much less in case of the Sun).
So the decay rate with the lifetime above (\ref{unutime}) is strongly
suppressed as compared to a corresponding annihilation rate for
the symmetric case.

\section{Conclusions}

In the present work it was shown that the positive result at 6.3
sigma C.L. obtained in DAMA/NaI and observed possible excesses in
cosmic gamma-radiation, positrons and antiprotons can be in
agreement within the framework of hypothesis of a   4th neutrino
mass hidden nearby half the Z-boson mass. The evident advantage of
this hypothesis is the minimal number of physical parameters. In
the simplest case it is only the mass of neutrino (to be compared
with minimally 5 parameters in the case of SUSY dark matter).

But even in this simplest case we have revealed the complex model
dependence
on the galactic mass distribution and cosmic ray diffusion. We
have shown its compatibility within realistic values of Heavy
neutrino mass. The model may be naturally extended in the case of
in-homogeneous  (clumpy) galactic halo, new Heavy neutrino
Interactions related to its necessary stability, relic neutrino
asymmetry and it consequent unstability and decay.
In those  models there are room for better agreement between
underground and cosmic rays signals.

The lightest  neutrino masses ($\sim 50$ GeV) might be searched
inside the old LEP data regarding electron pair annihilations into
one photon with missing energy \cite{accelerators}; the largest ones
($\sim 57-75$ GeV) might be
discovered by near future LHC search of invisible Higgs boson
decay\cite{Higgs}.
Additional satellites and antimatter search in Space might
define the exact parameter range for this extension of the
lepton sector, whose existence might be tested also in the
search for a novel pair of 4th quark family\cite{4quark}.

\section{Acknowledgment}

We are grateful to R.Bernabei, P.Belli and D.Prosperi for important
discussions.
K.B. thanks Universita' di Roma "Tor Vergata", INFN section Roma2 and
Universita di Roma' "La Sapienza" for their support and hospitality and
K.I. Shibaev for the help. M.Kh. is grateful to Abdus Salam ICTP (Trieste)
for hospitality.
This work was supported in part
%{\bf I think in part}
by grant RFBR 02-02-17490 and by the Federal
Program of the Russian Ministry of Industry,
Science and Technology 40.022.1.1.1106.


\begin{thebibliography}{XX}

\bibitem{DAMA-review}
R. Bernabei et al., Riv. Nuovo Cim. {\bf 26} n.1 (2003) 1;
astro-ph/0307403 and references therein.

\bibitem{Fargion}
D. Fargion, M.Yu. Khlopov, R.V. Konoplich and R. Mignani, JETP Lett. {\bf
68} (1998), 685;\\
D. Fargion, R. Konoplich, M. Grossi, M. Khlopov, astro-ph/9902327;\\
D. Fargion, et al., JETP Lett. {\bf 69} (1999), 434 ;
astro-ph-9903086.

\bibitem{BK} K.M. Belotsky, M.Yu. Khlopov, Gravitation and Cosmology
Suppl. {\bf 8} (2002), 112.\\
K.M. Belotsky, M.Yu. Khlopov, "Non-dominating heavy neutrino dark matter",
Proceedings of XVth conference "Physical cosmology" in Blois (2003).
%
%
%\bibitem[slow]{slow} K. Belotsky et al., Phys. Rev. {\bf D68} (2003),
%054027.

\bibitem{Okun} M. Maltoni et al., Phys. Lett. {\bf B476} (2000),
107;\\
V.A. Ilyin et al., Phys. Lett. {\bf B503} (2001), 126;\\
V.A. Novikov et al., Phys. Lett. {\bf B529} (2002) 111; JETP Lett. {\bf
76} (2002), 119.
%L.B. Okun, M.I. Vysotsky {\tt arXiv:hep-ph/0111028,hep-ph/0203132}.
%\bibitem[okun1]{okun1} S.S. Bulanov et al., Phys. Atom. Nucl. {\bf 66}
%(2003) 2169, and references therein.

\bibitem{Foot} R. Foot, Phys. Rev. {\bf D69} (2004), 036001;
[hep-ph/0308254];
R. Foot, astro-ph/0403043.

\bibitem{Z-Far} D. Fargion, B. Mele, A. Salis,  Astrophys. J. {\bf
517} (1999), 725.

\bibitem{Khlopov} V.K. Dubrovich, D. Fargion and M.Yu. Khlopov,
Astropart. Phys. {\bf 22} (2004), 183.

\bibitem{Datta} A. Datta, D. Fargion, B. Mele, hep-ph/0410176.

\bibitem{SM-0} I. Moskalenko, A. Strong, A.J. {\bf 493} (1998), 694;
A. Strong, I. Moskalenko, A.J. {\bf 509} (1998), 212.

\bibitem{SM-1} A. Strong, I. Moskalenko, O. Reimer, A.J. {\bf 537}
(2000), 763; astro-ph/9811296.

\bibitem{SM-2} A.Strong, I. Moskalenko, O. Reimer, M. Potgieter,
A.J. {\bf 565} (2002), 280; astro-ph/0106567.

\bibitem{Belli}
P. Belli et.al., Phys. Rev. {\bf D66} (2002), 043503.

\bibitem{Navarro} J.F. Navarro, C.S. Frenk, S.D.M. White, A.J.
{\bf 462}
(1996), 563.

\bibitem{SM-3} A. Strong, I. Moskalenko, O. Reimer,
astro-ph/0306345.

\bibitem{neutralino-gamma}
D. Els\"aesser, K. Mannheim, submitted to Physical Review Letters;
astro-ph/0405235.

\bibitem{solar modulation} L.J. Gleeson, W.I. Axford,
A.J. {\bf 154} (1968), 1011.

\bibitem{Casadei} D. Casadei, V. Bindi, astro-ph/0302307.

\bibitem{Ginzburg} V.S. Berezinskiy, S.V. Bulanov, S.V. Dogiel,
V.L. Ginzburg, V.S. Ptuskin, Astrophysics of Cosmic Rays
(Amsterdam: North Holand, 1990)

\bibitem{Turner}
M. Kamionkowski and M. S. Turner, Phys. Rev. {\bf D 43} (1991), 1774.


\bibitem{SM-4} I. Moskalenko, A. Strong, Phys. Rev. {\bf D} (1999);
astro-ph/9905283;\\
I. Moskalenko, A. Strong, Proceedings of 26th ICRC (Salt Lake City, 1999);
astro-ph/9906230.

\bibitem{reacceleration}  A.Strong, I.Moskalenko,
Proceedings of 3d INTEGRAL Workshop "The Extreme Universe",
14-18 Sep. 1998, Taormina, Italy; astro-ph/9811221.

\bibitem{HEAT} M.A. DuVernois et.al., A.J. {\bf 559} (2001),
296-303.

\bibitem{BESS} S. Orito et.al., Phys. Rev. Lett. {\bf 84} (2000),
1078; astro-ph/9906426.

\bibitem{BESS-all} T. Maeno et.al., Astropart. Phys. {\bf 16}
(2001), 121-128; astro-ph/0010381.

\bibitem{4thN-CR} R.V. Konoplich, M.Yu. Khlopov, Yad.Fiz.
{\bf 57} (1994), 452;\\
Yu.A. Golubkov, R.V. Konoplich, Yad.Fiz. {\bf 61} (1998), 675;
Phys.Atom.Nucl. {\bf 61} (1998), 602
%ALL OTHER APPROPRIATE REFERENCES ABSENT IN \cite{Fargion}

\bibitem{Berezinsky} V. Berezinsky, V. Dokuchaev, Yu.
Eroshenko, astro-ph/0301551.

\bibitem{ZKKC} Ya.B. Zeldovich, M.Yu. Khlopov, A.A. Klypin,
V.M. Chechetkin, Yad.Fiz. {\bf 31} (1980), 1286.
%

\bibitem{Shibaev}
M.Yu. Khlopov and K.I. Shibaev,
%``New Physics From Superstring Phenomenology''
Grav.\& Cosmol. Suppl.  {\bf 8 N1} (2002), 45.

\bibitem{Sakharov enhancement}
K.M. Belotsky, M.Yu. Khlopov, K.I. Shibaev,
Grav.\& Cosmol. Suppl.  {\bf 6} (2000), 140.

\bibitem{Sakharov}
A.D. Sakharov, ZhETF {\bf 18} (1948), 631; Reprinted in
Sov. Phys. Usp. {\bf 34} (1991), 375.

\bibitem{Dolgov}
A.D. Dolgov Phys.Rept. {\bf 370} (2002), 333; hep-ph/0202122.

\bibitem{accelerators}
D.Fargion, M.Khlopov, R.V.Konoplich and R.Mignani, Phys. Rev. {\bf D54}
(1996), 4684.

\bibitem{Higgs}
K.M.Belotsky, D. Fargion, M.Khlopov, R. Konoplich, K.Shibaev,  Phys.Rev.
{\bf D68} (2003), 054027; hep-ph/0210153.

\bibitem{4quark}
K.M.Belotsky, D. Fargion, M.Khlopov, R. Konoplich, M.Ryskin,
K.Shibaev,Grav.\& Cosmol. Suppl. (2005).
Invited talk at Cosmion 2004, Moscow-St.Petersburg-Paris(Meudon) 2-16,20-24
September 2004.

\end{thebibliography}
\end{document}